\documentclass[reprint,twocolumn,superscriptaddress,secnumarabic,amssymb, nobibnotes, aps,pra, floatfix]{revtex4-2}

\usepackage{amsmath}     
\usepackage{graphicx}    
\usepackage{verbatim}    
\usepackage{color}       
\usepackage{subfigure}   
\usepackage{hyperref}    
\usepackage{verbatim}    
\definecolor{ao}{rgb}{0.0, 0.5, 0.0}
\usepackage[normalem]{ulem}
\usepackage{lipsum}

\begin{document}


\title{Mechanism of intermetallic charge transfer and bond disproportionation \\ 
in BiNiO$_3$ and PbNiO$_3$ revealed by hard x-ray photoemission spectroscopy}

\author{Tatsuya~Yamaguchi}
\affiliation{Department of Physics and Electronics, Graduate School of Engineering, Osaka Metropolitan University 1-1 Gakuen-cho, Nakaku, Sakai, Osaka 599-8531, Japan.}

\author{Mizuki~Furo}
\affiliation{Department of Physics and Electronics, Graduate School of Engineering, Osaka Metropolitan University 1-1 Gakuen-cho, Nakaku, Sakai, Osaka 599-8531, Japan.}

\author{Yuki~Sakai}
\affiliation{Kanagawa Institute of Industrial Science and Technology (KISTEC), Ebina 243-0435, Japan}
\affiliation{Materials and Structures Laboratory, Institute of Innovative Research, Tokyo Institute of Technology 4259 Nagatsuta, Midori, Yokohama 226-8501, Japan}

\author{Takumi~Nishikubo}
\affiliation{Kanagawa Institute of Industrial Science and Technology (KISTEC), Ebina 243-0435, Japan}
\affiliation{Materials and Structures Laboratory, Institute of Innovative Research, Tokyo Institute of Technology 4259 Nagatsuta, Midori, Yokohama 226-8501, Japan}

\author{Hajime~Hojo}
\affiliation{Department of Advanced Materials Science and Engineering, Faculty of Engineering Sciences, Kyushu University, 6-1 Kasuga-koen, Kasuga, Fukuoka 816-8580, Japan}

\author{Masaki~Azuma}
\affiliation{Materials and Structures Laboratory, Institute of Innovative Research, Tokyo Institute of Technology 4259 Nagatsuta, Midori, Yokohama 226-8501, Japan}
\affiliation{Kanagawa Institute of Industrial Science and Technology (KISTEC), Ebina 243-0435, Japan}

\author{Kengo~Oka}
\affiliation{Department of Applied Chemistry, Faculty of Science and Engineering, Chuo University, 1-13-27 Kasuga, Bunkyo-ku, Tokyo 112-8551, Japan.}
\affiliation{Department of Applied Chemistry, Faculty of Science and Engineering, Kindai University, 3-4-1 Kowakae, Higashi-Osaka, Osaka 577-8502, Japan.}

\author{Daisuke~Mori}
\affiliation{Department of Chemistry, Faculty of Science, Gakushuin University, 1-5-1 Mejiro, Toshima-ku, Tokyo 171-8588, Japan.}
\affiliation{Department of Applied Chemistry, Graduate School of Engineering, Mie University, Tsu, Mie 514-8507, Japan.}

\author{Yoshiyuki~Inaguma}
\affiliation{Department of Chemistry, Faculty of Science, Gakushuin University, 1-5-1 Mejiro, Toshima-ku, Tokyo 171-8588, Japan.}

\author{Masaichiro~Mizumaki}
\affiliation{Institute of Industrial Nanomaterials, Kumamoto University 2-39-1 Kurokami Chuo-ku Kumamoto 860-8555, Japan.}

\author{Kento~Yamamoto}
\affiliation{Department of Applied Physics, Waseda University, 3-4-1 Okubo, Shinjuku-ku, Tokyo 169-8555, Japan.}

 \author{Jan Kune\v{s}}
\affiliation{Institute of Solid State Physics, TU Wien, 1040 Vienna, Austria}
\affiliation{Department of Condensed Matter Physics, Faculty of
  Science, Masaryk University, Kotl\'a\v{r}sk\'a 2, 611 37 Brno,
  Czechia}

\author{Takashi~Mizokawa}
\affiliation{Department of Applied Physics, Waseda University, 3-4-1 Okubo, Shinjuku-ku, Tokyo 169-8555, Japan.}

\author{Atsushi~Hariki}
\affiliation{Department of Physics and Electronics, Graduate School of Engineering, Osaka Metropolitan University 1-1 Gakuen-cho, Nakaku, Sakai, Osaka 599-8531, Japan.}

\date{\today}

\begin{abstract}

Perovskites with Bi or Pb on the A site host a number of interesting and yet to be understood
phenomena such as negative thermal expansion in BiNiO$_3$.
We 
employ hard x-ray photoemission spectroscopy of Ni 2$p$ core level as well as valence band
to probe the electronic structure of BiNiO$_3$ and PbNiO$_3$. The experimental results
supported by theoretical calculations using dynamical mean-field theory reveal
essentially identical electronic structure of the Ni--O subsystem typical
of Ni$^{2+}$ charge-transfer insulators. The two materials are distinguished by filling
of the Bi(Pb)--O antibonding states in the vicinity of the Fermi level, which is responsible
for the Bi disproportionation in BiNiO$_3$ at ambient pressure and absence of similar behavior in 
PbNiO$_3$. The present experiments provide evidence for this conclusion by revealing
the presence/absence of Bi/Pb $6s$ states at the top of the valence band in the two materials.


\end{abstract}

\maketitle

\section{Introduction}

The A sites of typical AMO$_3$ perovskite transition-metal oxides (TMOs) are occupied by divalent alkaline earth metal or trivalent Y or rare-earth ions, which act as electron donors and form ionic bonds with oxygen.
Perovskites with Pb or Bi on the A site are different.
The Pb or Bi ions 
play a more active role thanks to their ability to form covalent bonds with oxygen.
Both Pb and Bi belong to so-called valence skippers, ions that prefer to adopt the 6$s^0$ (Pb$^{4+}$, Bi$^{5+}$) and 6$s^2$ (Pb$^{2+}$, Bi$^{3+}$) valence configurations while avoiding the 6$s^1$ (Pb$^{3+}$, Bi$^{4+}$) configuration. 
This behavior can be modelled with an attractive onsite interaction, negative $U$, within the 6$s$ shell~\cite{Matsuura2022}. The origin of the negative $U$ has been attributed to covalent bonding plus electron-lattice coupling~\cite{Anderson1975} or nonlinear nonlocal screening~\cite{Varma1988}. 

Electronic structure calculations~\cite{paul19,Hariki2021} have shown that valence skipping in perovskite TMOs is caused by Pb(Bi)--O hybridization, which leads to formation of a relatively narrow antibonding band. Its energy, determined by bonding-antibonding splitting, is sensitive to small changes of the Pb(Bi)--O bond lengths and therefore strongly coupled to the lattice~\cite{An2001}. Formal 6$s^0$ and 6$s^2$ valence configurations
correspond to a situation where the antibonding band is located above and below the Fermi level $E_F$, respectively.
The $6s^1$ or partially filled configurations would amount to pinning of the antibonding band at $E_F$. Such states/structures are often unstable towards small changes of the Pb(Bi)--O lengths, which cause the antibonding states to move away from $E_F$. This is accompanied by transfer of electrons/holes to other orbitals,~e.g., TM $3d$ ones. Alternatively, Pb(Bi)--O bond-length disproportionation splits the antibonding band away from $E_F$ and leads to redistribution of charge among inequivalent
Pb(Bi) ions.
While the negative $U$ picture also leads to valence skipping it does not account for the geometrical constraints and correlations between Pb(Bi)--O bond lengths due to the A sites sharing their oxygen neighbors and vice versa.
This affects possible ordering patterns of 6$s^2$ and 6$s^0$ sites on the lattice~\cite{paul19,Hariki2021} as well as dynamics of the 6$s$ electrons.

The Pb(Bi)--O hybridization picture can be probed by studying the Pb(Bi) $6s$ spectra.
To this end, we have measured the valence and core-level x-ray photoemission spectroscopy (XPS) spectra of closely related PbNiO$_3$ and BiNiO$_3$ 
and analyzed them with the help of electronic structure calculations using local-density approximation (LDA) combined with dynamical mean-field theory (DMFT) method. We employ hard x-ray photoemission spectroscopy (HAXPES) which has large photoionization cross sections (CSs) for the Pb(Bi) 6$s$ orbitals.

At ambient pressure BiNiO$_3$ is a good insulator~\cite{Ishiwata05} while PbNiO$_3$ shows a semiconducting behavior~\cite{Inaguma11}.
 BiNiO$_3$ undergoes a pressure-driven metal-insulator transition (MIT) at the critical pressure 3.5~GPa~\cite{Azuma11,Azuma07}, which has been identified as a new type of MIT driven by electron transfer from Bi to Ni and interpreted using the negative $U$ picture~\cite{Naka16,Kojima16}. 
 At ambient pressure, BiNiO$_3$ is an insulator with two distinct Bi sites and formal valence Bi$^{3+}_{0.5}$Bi$^{5+}_{0.5}$Ni$^{2+}$O$_3$~\cite{Ishiwata02}. At high pressure, it is a metal with formal valence Bi$^{3+}$Ni$^{3+}$O$_3$~\cite{Azuma11,Azuma07}. PbNiO$_3$ takes formal valence Pb$^{4+}$Ni$^{2+}$O$_3$.

\section{Methods}

\subsection{Experimental method}

Pollycrystalline samples of 
BiNiO$_3$ and PbNiO$_3$
were prepared under high pressure~\cite{Inaguma11}. HAXPES measurements for valence band and Ni 2$p$ core-level were performed at 300 K with photon energy of 
7939.8~eV at the undulator beamlines BL47XU of SPring-8. For these measurements, hemispherical photoelectron analyzers (SES2002/R-4000) were used. The polycrystalline samples were fractured {\it in situ} for the XPS measurements. The binding energies were calibrated using Au 4f$_{7/2}$ peak (84.0~eV) and Fermi edge of gold reference samples. The total energy resolution was 280~meV.

\subsection{Computational method}

\begin{figure}
    \includegraphics[width=0.96\columnwidth]{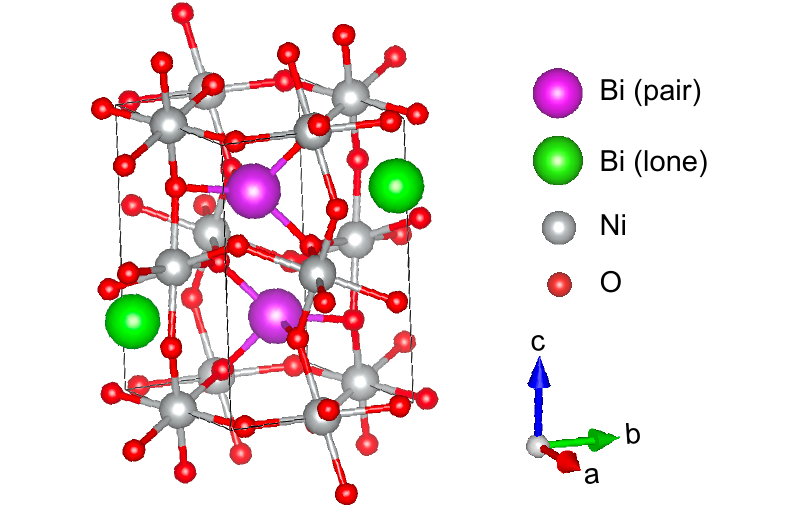}
    \caption{Crystal structure of BiNiO$_3$~\cite{Ishiwata02} visualized by VESTA~\cite{vesta}.}
    \label{fig_struct}
\end{figure}

The valence-band and core-level XPS spectra are simulated using the
LDA+DMFT method~\cite{metzner89,georges96,kotliar06}.
The computation starts with a standard LDA calculation using Wien2k~\cite{wien2k} for the experimental crystal structures of PbNiO$_3$ ($Pbn2_1$ space group)~\cite{sakai19} and BiNiO$_3$ ($P\overline{1}$ space group)~\cite{Ishiwata02}.
In BiNiO$_3$, two inequivalent Bi sites~\cite{Ishiwata02} are present as shown in Fig.~\ref{fig_struct}.
The two Bi sites, with the shortest Bi--O bonds of 2.21~\AA~and 2.03~\AA, are referred to as the ``lone'' and ``pair'' Bi ions hereafter.

Next, the LDA bands are mapped onto a tight-binding model spanning the Bi (Pb) 5$d$, 6$s$, 6$p$, Ni 3$d$, 4$s$, O 2$s$ and 2$p$ bands~\cite{wien2wannier,wannier90}. 
Although the low-energy electronic structures of the two compounds are governed by the Ni 3$d$, O 2$p$ and Bi (Pb) 6$s$ states, the other orbitals can be observed in the HAXPES spectra due to hybridization.
The tight-binding model is augmented with a local electron-electron interaction within the Ni 3$d$ shell. We employ Hubbard $U$ and Hund's $J$ parameters of ($U$, $J$) = (7.0~eV, 1.1~eV)~\cite{Anisimov91}. The double-counting correction $\mu_{\rm dc}$~\cite{kotliar06,karolak10} is set to reproduce the experimental valence-band and core-level spectra as in Refs.~\cite{hariki17,Higashi21,Hariki22}. 
\textcolor{black}{The details can be found in the Supplementary Material (SM)~\cite{sm}.}
The strong-coupling continuous-time quantum Monte Carlo impurity solver~\cite{werner06,boehnke11,hafermann12}
of the Anderson impurity model (AIM) is used to compute Ni 3$d$ self-energies $\Sigma(i\omega_n)$. The valence-band spectra and hybridization densities $\Delta(\omega)$ on the real frequency axis are calculated from the local self-energy $\Sigma(\omega)$ analytically continued to the real-frequency axis using maximum entropy method~\cite{wang09,jarrell96}. 

The Ni 2$p$ core-level XPS spectra are calculated 
using a configuration-interaction impurity solver~\cite{hariki17,Ghiasi2019,winder20}
from the DMFT AIM 
augmented with the Ni 2$p$ core orbitals and their interaction with the valence electrons. 
 The hybridization densities $\Delta(\omega)$ are represented by 25 discrete levels (per spin and orbital). 


\section{Results and Discussion}

\begin{figure}
    \includegraphics[width=0.99\columnwidth]{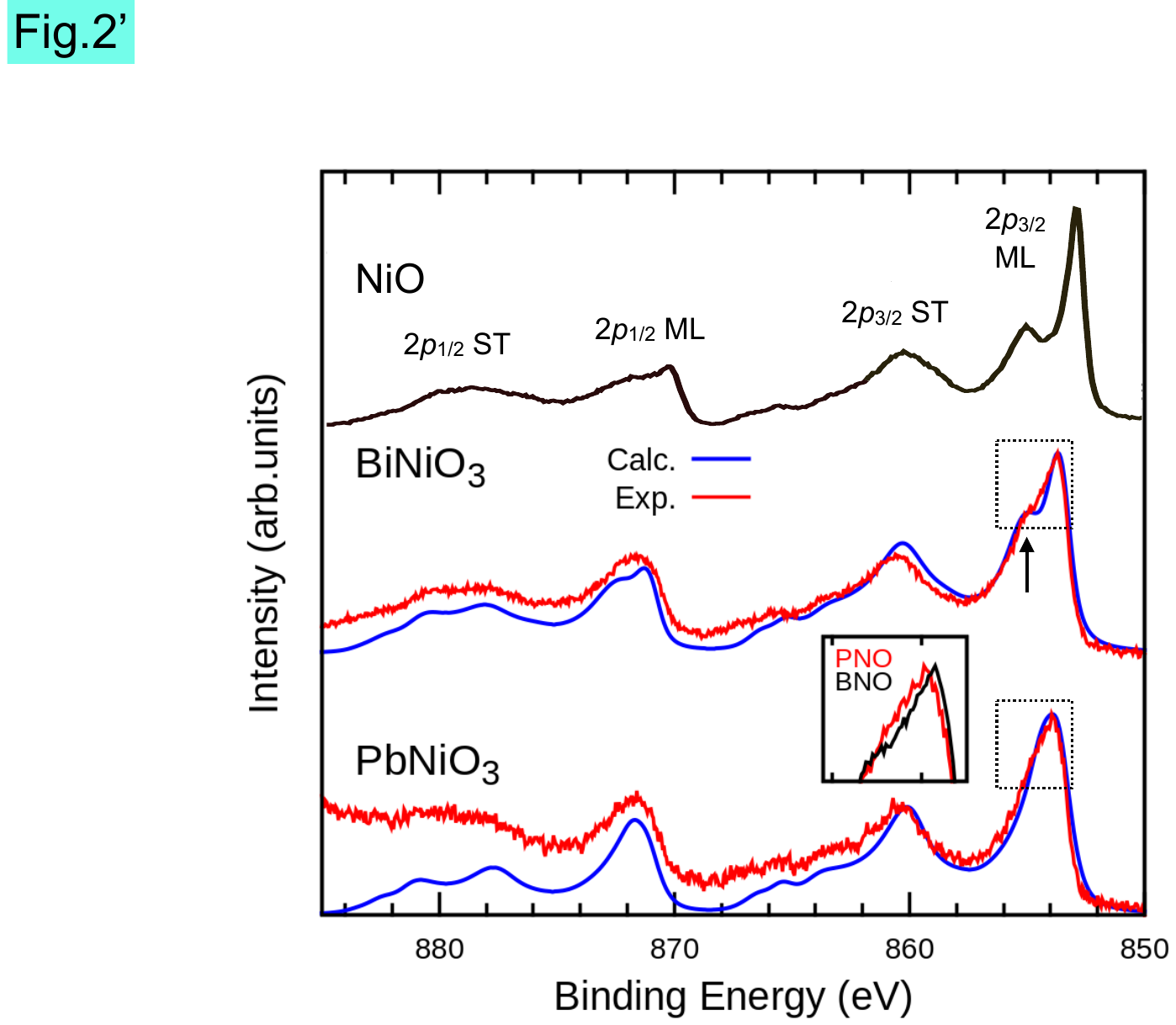}
    \caption{\textcolor{black}{The Ni 2$p$ HAXPES experimental data (red) and the LDA+DMFT AIM calculations (blue) for BiNiO$_3$ in the AFI phase (middle) and PbNiO$_3$ (bottom) in the PMI phase. An arrow indicates the shoulder feature mentioned in the main text. The inset displays the magnified view of the Ni $2p_{3/2}$ ML in the experimental data for PbNiO$_3$ (PNO, red) and BiNiO$_3$ (BNO, black). The spectral broadening is considered using Gaussian of 300~meV and Lorentian of 300~meV (HWHM) in the calculated results. The Ni 2$p$ XPS data of NiO~\cite{taguchi08} is shown for comparison (top). The Ni 2$p_{3/2}$ (2$p_{1/2}$) main line (ML) and satellite (ST) are indicated.} }
    \label{fig_xps}
\end{figure}

\begin{figure}
    \includegraphics[width=0.99\columnwidth]{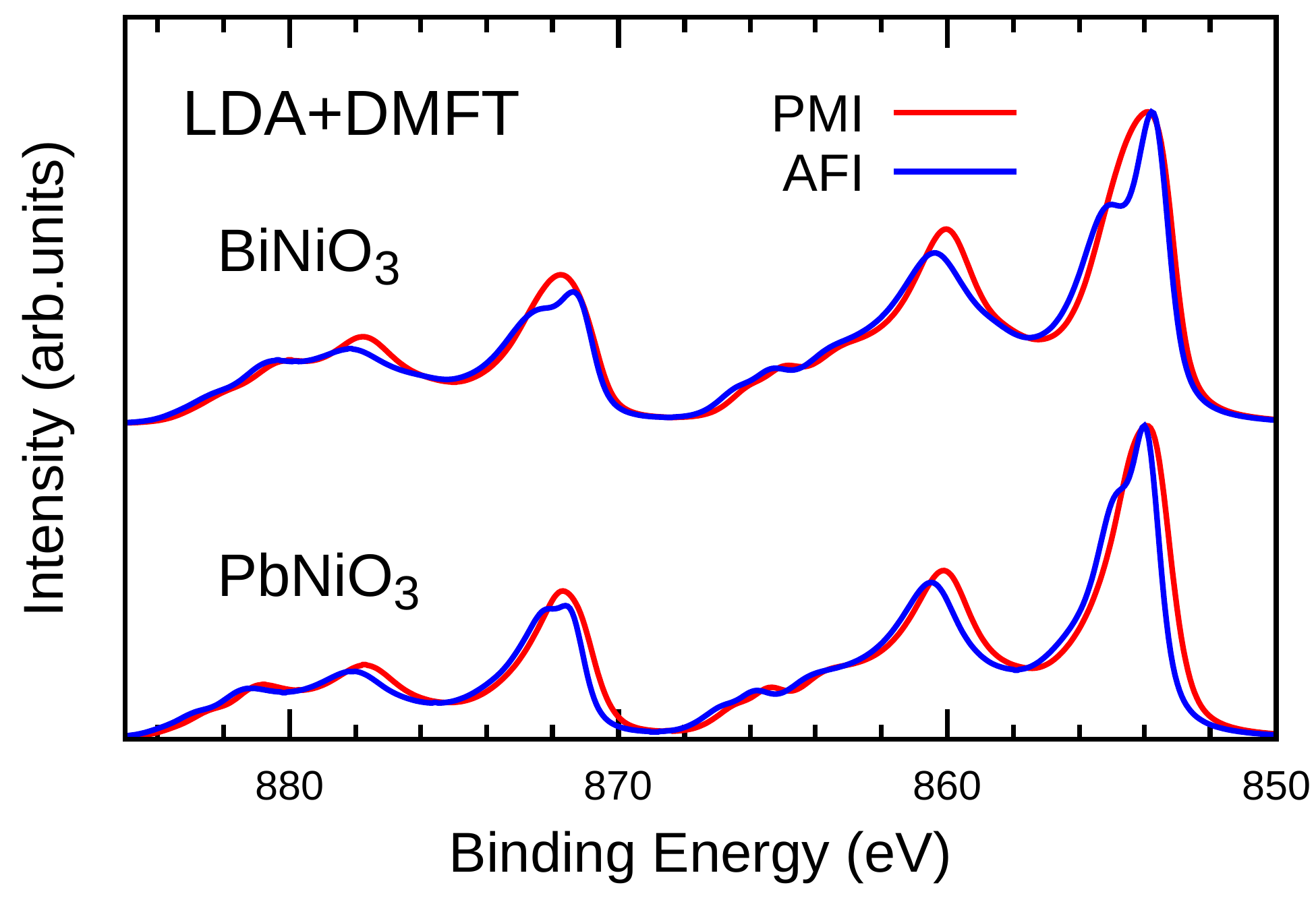}
    \caption{The LDA+DMFT Ni 2$p$ XPS spectra calculated for BiNiO$_3$ (top) and PbNiO$_3$ (bottom) in the paramagnetic insulating (red) and \textcolor{black}{antiferromagnetic insulating} (blue) phases. }
    \label{fig_xps_pmaf}
\end{figure}

\subsection{Core-level XPS}
We start our discussion with Ni 2$p$ core-level XPS spectra.
Figure~\ref{fig_xps} shows the Ni 2$p$ XPS spectra of BiNiO$_3$ and PbNiO$_3$ measured at room temperature. 
Overall spectral shapes are similar to those of typical Ni$^{2+}$ oxides, such as NiO~\cite{taguchi08} and La$_2$NiO$_4$~\cite{Eisaki92}.
Both the 2$p_{1/2}$ (884--868~eV) and 2$p_{3/2}$ (868--852~eV) contributions consist of a main line (ML) and a satellite (ST).
The Ni 2$p_{3/2}$ ML in BiNiO$_3$ exhibits a shoulder on the higher-binding energy side,
while the shoulder is less pronounced and the ML is more symmetric in PbNiO$_3$.
Given that the N\'eel temperatures of 300~K in BiNiO$_3$ and 225~K in PbNiO$_3$~\cite{Ishiwata02, Inaguma11},
we show the simulations for the antiferromagnetic insulating (AFI) phase for BiNiO$_3$ and the paramagnetic insulating (PMI) phase for PbNiO$_3$ in Fig.~\ref{fig_xps}. Figure~\ref{fig_xps_pmaf} shows that
the spectra calculated within the same phase (i.e.~AFI or PMI) are very similar in the two compounds, suggesting essentially identical electronic structures of Ni–O subsystems in BiNiO$_3$ and PbNiO$_3$.

We briefly discuss the physical origin of the observed core-level spectra.
Sudden creation of a localized core hole triggers a dynamical charge response
in the XPS final states, which amounts to charge transfer (CT) from surrounding ligands as well as distant Ni ions to the excited Ni site, conventionally referred to as CT screening. 
The two components in the ML originate from two CT screening processes:~one from nearest-neighboring ligands and the other, at lower binding energy, from distant Ni ions, often called local and nonlocal screening, respectively~\cite{veenendaal93,hariki13b,taguchi08}. 
The nonlocal CT screening in Ni$^{2+}$ oxides~\cite{hariki13b,hariki17} 
is less efficient in PMI than in AFI due to Pauli blocking, which 
explains the suppression of low-energy peak in PMI.

\begin{figure}
    \includegraphics[width=0.99\columnwidth]{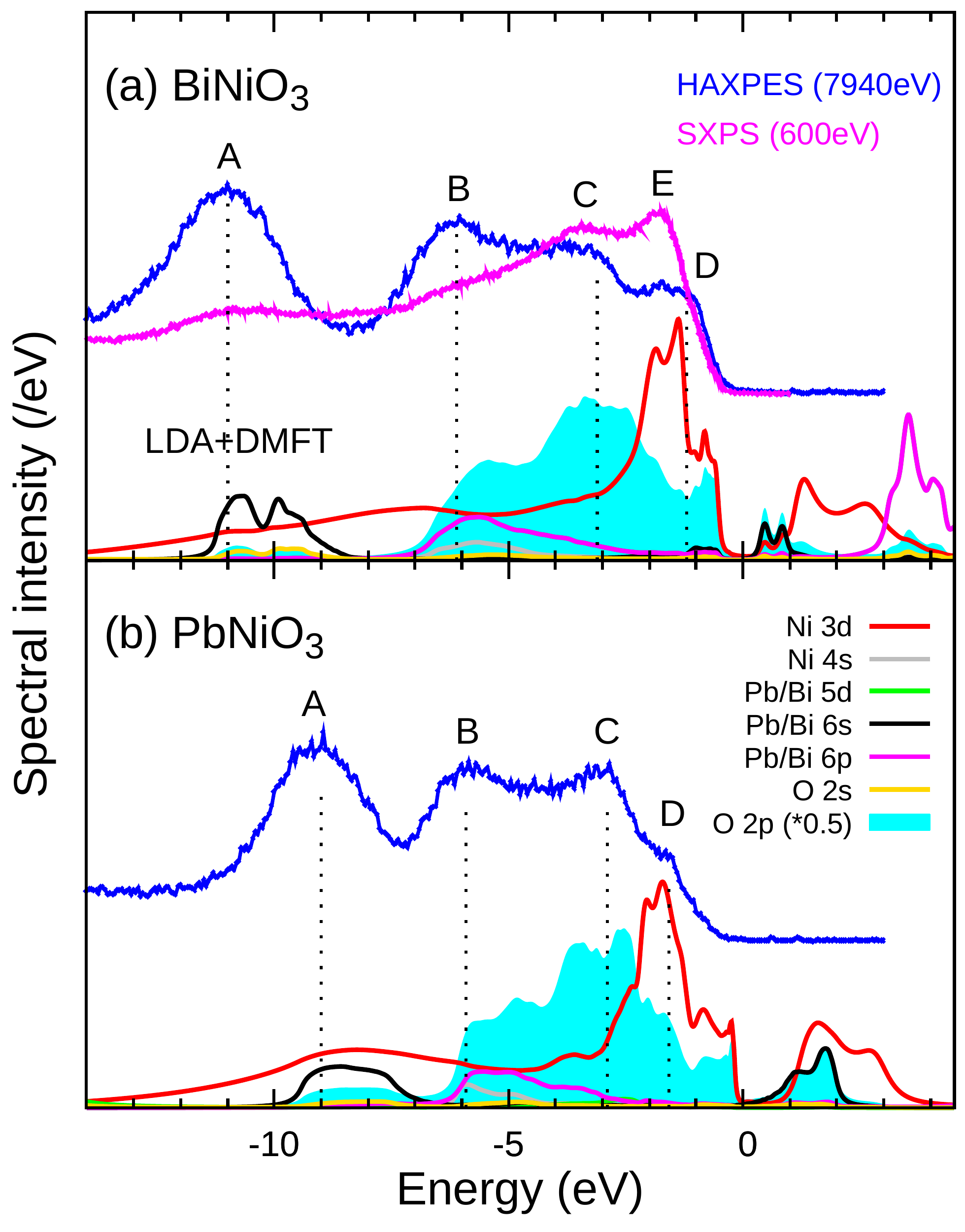}
    \caption{The LDA+DMFT valence-band spectral intensities of (a) BiNiO$_3$ in the \textcolor{black}{antiferromagnetic} phase and (b) PbNiO$_3$ in the paramagnetic phase. 
    The experimental valence-band spectra (blue) of BiNiO$_3$~\cite{Wadati07, Nishikubo19} and PbNiO$_3$ are shown for comparison.
    }
    \label{fig_dos}
\end{figure}

\subsection{Valence-band XPS}



Figure~\ref{fig_dos} shows the experimental and LDA+DMFT valence-band spectra of BiNiO$_3$~\cite{Nishikubo19} and PbNiO$_3$. 
Soft x-ray XPS (SXPS) spectrum of Bi$_{0.95}$La$_{0.05}$NiO$_3$ measured at photon energy of 600~eV~\cite{Wadati07} is shown for comparison. 
The difference between the spectra obtained with hard and soft x rays reflects the different photon-energy dependencies of the photoionization CSs
for the constituent orbitals. The empirical CS values summarized in 
Table~\ref{Tab_cs_value} show
that while Ni 3$d$ orbitals have large CS at 500~eV, at 7940~eV
the Bi (Pb) CSs are dominant.
This makes the hard x rays well suited to study the Bi and Pb $6s$ states.

We identify features labeled as $A$, $B$, $C$, $D$ in the HAXPES spectra
of BiNiO$_3$ and PbNiO$_3$. In addition, a feature $E$, enhanced in soft x-ray spectra, is attributed to the Ni 3$d$ orbital contributions and matches well the position of the Ni 3$d$ peak in the LDA+DMFT result. 
The peak $A$ corresponds to Bi (Pb) 6$s$ states--bonding state in the language of Bi(Pb)--O hybridization.
\textcolor{black}{On the other hand, peak $D$ gains its weight in the HAXPES spectrum of BiNiO$_3$, peak $D$ of BiNiO$_3$ can be assigned to Bi 6$s$ or 6$p$.}

\begin{figure}
    \includegraphics[width=0.99\columnwidth]{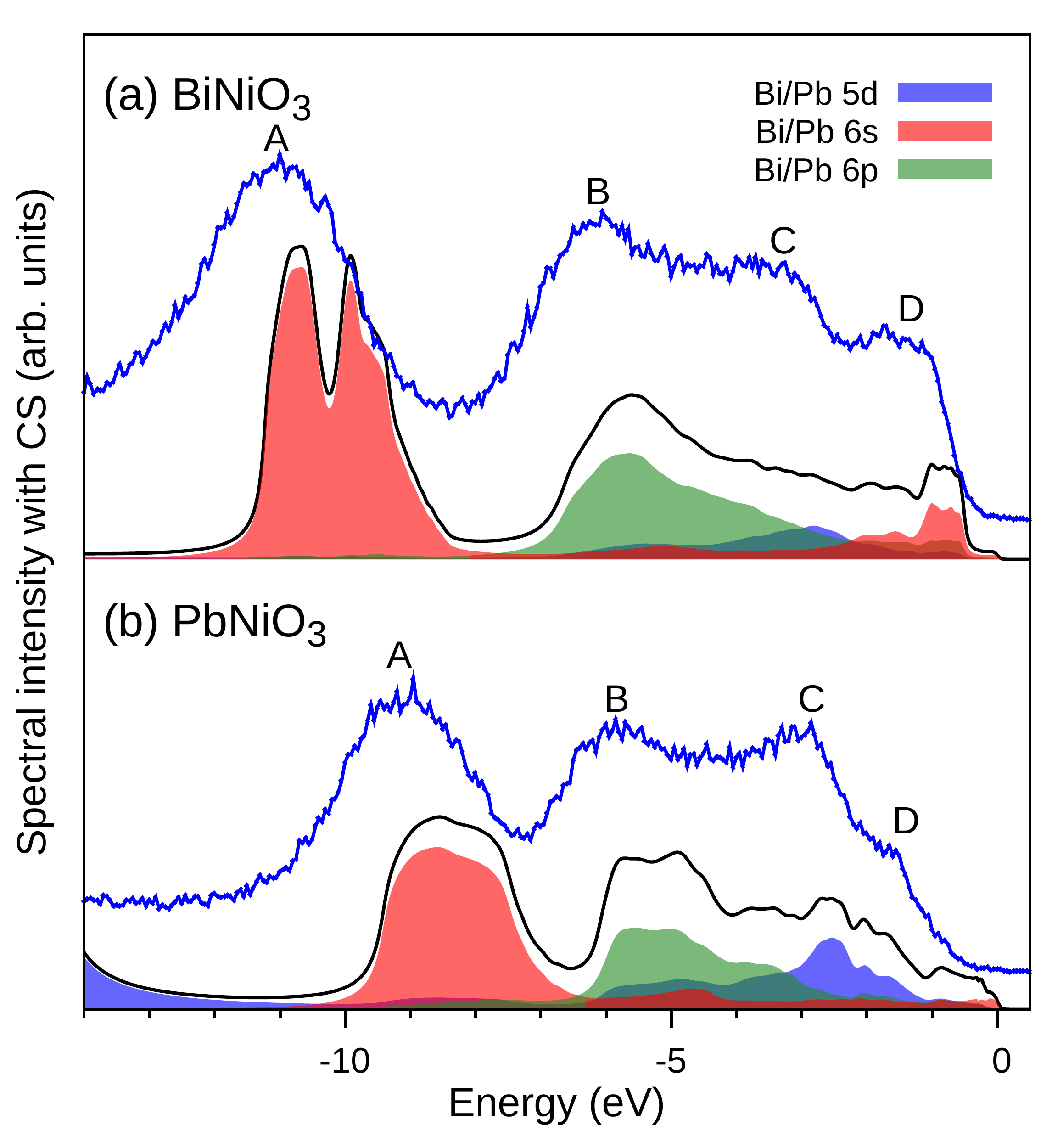}
    \caption{The LDA+DMFT valence-band spectral intensities of (a) BiNiO$_3$ in the antiferromagnetic phase and (b) PbNiO$_3$ in the paramagnetic phase weighted with the CSs listed in Table.~\ref{Tab_cs_value}. 
     The Fermi-Dirac function at 300~K is included in the  theoretical spectra.
    The filled curves show the contributions of the Bi (Pb) 5$d$, 6$s$ and 6$p$ states.
    The experimental HAXPES data (blue) of BiNiO$_3$~\cite{Nishikubo19} and PbNiO$_3$ are shown for comparison.
    }
    \label{fig_dos_cs}
\end{figure}

In order to identify the origin of features $B$, $C$, $D$, we weight the LDA+DMFT valence-band intensities with the CSs at 7940~eV in Fig.~\ref{fig_dos_cs}. 
\textcolor{black}{An underestimate ($\sim 0.5$~eV) of the binding energies of the feature $A$ is due to uncertainty in the adopted approximation for the exchange-correlation potential~\cite{Hao_2014}.}
Although the Bi (Pb) 5$d$ and 6$p$ states are positioned far below and above $E_F$, see SM~\cite{sm}, 
their CSs are extremely large in the hard x-ray regime, Table~\ref{Tab_cs_value}.
This together with a tiny hybridization with the valence band makes them apparent in the HAXPES spectra in the valence region. 
The simulated spectra in Fig.~\ref{fig_dos_cs} show that the Bi or Pb 6$p$ states are important for the feature $B$, while the $5d$ states give rise to the feature $C$.
The shallower 
Pb 5$d$ states could explain the enhanced intensity of the feature $C$ in PbNiO$_3$ compared to that in BiNiO$_3$.

      \begin{table}
      \centering
        \begin{ruledtabular}
        \begin{tabular}{ c | c  c  c | c  c}
            \quad    & {\quad} & 8000~eV  & {\quad} &  \multicolumn{2}{c}{500~eV}   \\
            \hline
            \quad    & $\beta$ &  $\sigma$~{\lbrack}b{\rbrack} & $d\sigma/d\Omega$~{\lbrack}b/sr{\rbrack} & $\sigma$~ {\lbrack}b{\rbrack} &  $d\sigma/d\Omega$~ {\lbrack}b/sr{\rbrack} \\
            \hline
            Ni 3$d$  &  0.322  &   0.741  &   0.078  &  19.969  &   1.589  \\
            Ni 4$s$  &  1.960  &   6.855  &   1.615  &   2.569  &   0.204  \\
            Bi 5$d$  &  1.198  &  68.690  &  12.014  &  14.282  &   1.137  \\
            Bi 6$s$  &  1.890  &  25.050  &   5.761  &   2.658  &   0.212  \\
            Bi 6$p$  &  1.731  &  14.340  &   3.116  &   1.569  &   0.125  \\
            Pb 5$d$  &  1.184  &  59.350  &  10.317  &  13.116  &   1.044  \\
            Pb 6$s$  &  1.898  &  21.495  &   4.957  &   2.338  &   0.186  \\
            Pb 6$p$  &  1.653  &  13.475  &   2.845  &   0.983  &   0.078  \\
            O 2$s$   &  1.923  &   6.795  &   1.581  &  11.510  &   0.916  \\
            O 2$p$   &  0.053  &   0.080  &   0.007  &   2.214  &   0.176  \\  
        \end{tabular}
        \caption{Subshell photoionization cross section $\sigma$ at 8000~eV (hard X-ray) and 500~eV (soft X-ray) for relevant orbitals~\cite{Trzhaskovskaya18, Trzhaskovskaya01, Trzhaskovskaya02}. The differential cross section $d\sigma/d\Omega$ at 8000~eV is calculated taking the dipole parameter $\beta$ for the angular distribution into account. To simulate the experimental geometry in the previous SXPS study in Ref.~\cite{Wadati07}, $\beta$ is not considered for $d\sigma/d\Omega$ at 500~eV.}
        \label{Tab_cs_value}
        \end{ruledtabular}
        \end{table}

\begin{figure}
    \includegraphics[width=0.99\columnwidth]{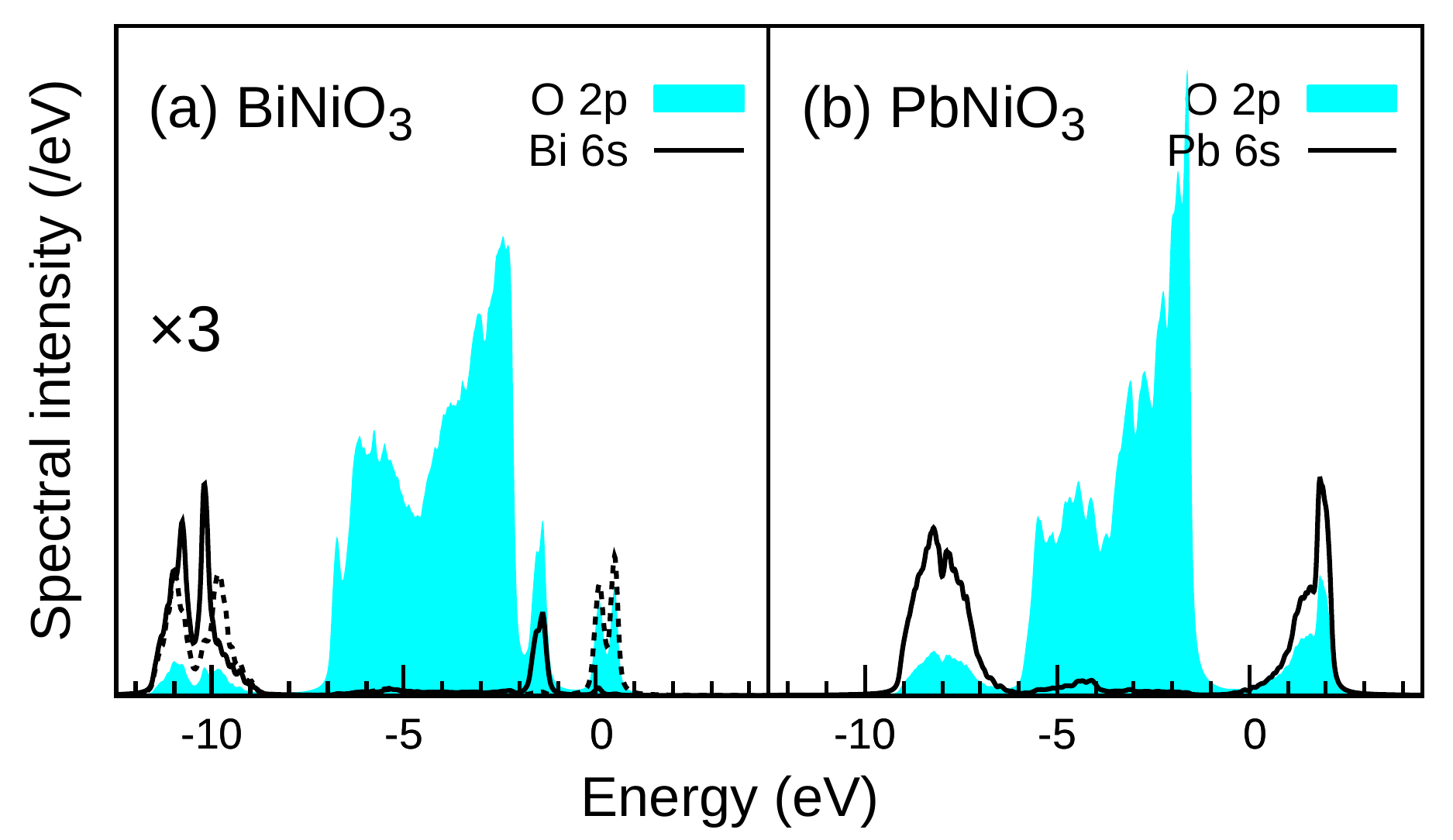}
    \caption{The valence-band spectra in (a) BiO$_3$ and (b) PbO$_3$ sublattice computed by removing the Ni 3$d$ orbitals in the LDA+DMFT valence-band models in Figs.~\ref{fig_dos}ab.
    The Bi and Pb $6s$ spectral weights are magnified by the factor of 3. The solid (dashed) line in BiNiO$_3$ represents lone (pair) Bi site contribution.}
    \label{fig_dos_demo}
\end{figure}
The main qualitative difference between the HAXPES spectra of BiNiO$_3$ and PbNiO$_3$ is
the enhanced spectral weight at the top of the valence band.
The analysis of Fig.~\ref{fig_dos_cs} links it to the presence of Bi $6s$ orbital character, while Pb $6s$ states are 
missing at the valence-band top in PbNiO$_3$.
To isolate the physics of Bi(Pb) $6s$--O 2$p$ bonding, we present, in Fig.~\ref{fig_dos_demo}, the Bi(Pb) 6$s$ and O 2$p$ spectral densities obtained with the Ni--O and Ni--Bi(Pb) hopping artificially 
switched off in the LDA+DMFT valence-band models in Fig.~\ref{fig_dos}.
Besides the quantitative differences consisting in shallower $6s$ states
for lighter Pb and the resulting somewhat stronger covalent character of the Pb--O
bond, there is a qualitative difference reflecting the disproportionation
of the Bi sites. This leads to splitting of the antibonding band into
states associated with the lone Bi site at lower and pair Bi sites at higher energies,
consistent with the previous GGA+$U$ and DMFT studies~\cite{paul19,Leonov19}.
No splitting of the antibonding band takes place in PbNiO$_3$. This picture,
necessarily with broader and richer structures in the $6s$ spectra, is preserved
when Ni is fully included. The CT gap of Ni--O subsystem overlaps with 
lone-pair antibonding gap in BiNiO$_3$ and O $2p$-antibonding gap in PbNiO$_3$ in Fig.~\ref{fig_dos}.

\section{Conclusions}
We have presented a comparative study of two A-site active nickelates 
BiNiO$_3$ and PbNiO$_3$. The valence and Ni 2$p$ core-level photoemission experiments
supported by LDA+DMFT calculations reveal 
essentially
identical electronic structures
of Ni--O subsystems in the two compounds, characteristic of Ni$^{2+}$ 
in charge-transfer systems such as NiO~\cite{Sawatzky1984,Fujimori1984,Kunes2007a}.
The main difference between the two compounds
is derived from
the Pb(Bi) $6s$ orbitals.
The calculations find strong Pb(Bi)--O hybridization leading to bonding-antibonding splitting of almost 10~eV, with the antibonding states located within the charge-transfer gap. In PbNiO$_3$, electron counting places the Fermi level below the antibonding band,
while in BiNiO$_3$ the Fermi level falls in the middle of the antibonding band. 
This situation in the latter is resolved by disproportionation of the Bi--O bonds, which gives rise to two distinct Bi sites, with antibonding states above and below the Fermi level.

Our main experimental result is the observation of the Bi--O antibonding states at the
top of the valence band, which confirms the above mechanism of Bi disproportionation.
We conclude our presentation by commenting on the relationship of the present
description to the negative $U$ model of valence skippers. The negative $U$ model qualitatively 
explains presence (absence) of A-site disproportionation in BiNiO$_3$ (PbNiO$_3$). It can be viewed as an effective model for the Pb(Bi)--O antibonding states
obtained by the process of integrating out the O states. However, in this process 
also nonlocal interaction terms are expected to appear, e.g., due to O site having 
several Pb(Bi) neighbors. Such terms determine the pattern of disproportionated Bi
ions on the lattice. 

A question also arises about the negative $U$ treatment of possible superconductivity 
in materials with valence skippers. Strong covalent bonds in systems with antibonding states at the Fermi level result in a strong electron-phonon coupling~\cite{An2001}. As in other such materials,
superconductivity is competing with a structural instability. 
The capability of the negative $U$ model to capture this competition 
appears questionable and we find that the full treatment
of Pb(Bi)--O bonding without the effective
$6s$ interaction $U$ preferable.

\begin{acknowledgments}

This work was supported by JSPS KAKENHI Grants No.~21K13884, No.~21H01003, No.~23K03324, No.~23H03816, No.~23H03817 (A.H.), JP19H0562 and JSTCREST (JPMJCR22O1) (M.A.), and by the Project No. CZ.02.01.01/00/22\_008/0004572 of the Programme Johannes Amos Commenius (J.K.). The synchrotron-radiation experiments were performed at SPring-8 with the approval of the Japan Synchrotron Radiation Research Institute (2017B1721). The authors would like to thank Dr.~Shigenori Ueda for the contribution to HAXPES measurements at BL15XU, SPring8, in the early stage of this work.

\end{acknowledgments}

\bibliography{main}

\end{document}